\documentclass[journal]{IEEEtran}
\usepackage{amsmath,amssymb}
\usepackage{makecell}
\usepackage{color}
\usepackage{graphicx,subfig,graphics,float}
\graphicspath{{figures/}}
\usepackage{enumerate}
\usepackage{multirow}
\usepackage{dsfont}
\usepackage{cleveref}
\usepackage{easybmat}

\usepackage{tkz-euclide}
\usepackage{tikz-3dplot}
\usepackage{pgfplots}
\pgfplotsset{compat=1.11}
\usepgfplotslibrary{fillbetween}
\usepackage{tikz}
\usetikzlibrary{shapes,arrows,calc,positioning,automata,intersections,plotmarks,quotes,angles}
\usetikzlibrary{positioning}
\usetikzlibrary{decorations.pathmorphing} 
\usetikzlibrary{matrix,fit} 
\tikzstyle{block} = [draw, rectangle, thick, text centered, minimum height=0.7cm, minimum width=1cm]
\tikzstyle{sum} = [draw, thick, circle, node distance=1.5cm] 
\tikzstyle{guide} = [coordinate] 
\tikzstyle{smallblock} = [draw, rectangle, minimum height=0.7cm, minimum width=3cm]

\DeclareMathOperator*{\diag}{\mathrm{diag}}

\DeclareMathOperator*{\E}{\mathrm{\bf E}}
\DeclareMathOperator*{\Cov}{\mathrm{\bf Cov}}

\DeclareMathOperator*{\Zset}{\mathbb{Z}}

\newtheorem{thm}{Theorem}
\newtheorem{lma}{Lemma}
\newtheorem{rmk}{Remark}

\newtheorem{defn}{Definition}
\newtheorem{assump}{Assumption}

\newtheorem{corollary}{Corollary}

\newcommand{\vt}{\tilde{V}}
\newcommand{\ut}{\tilde{U}}
\newcommand{\mt}{\tilde{M}}
\newcommand{\nt}{\tilde{N}}
\newcommand{\mi}{M_{in}}
\newcommand{\mo}{M_{out}}
\newcommand{\nto}{\tilde{N}_{out}}
\newcommand{\mgi}{M_{\Gamma in}}

\newcommand{\mgit}{M_{\Gamma in,\bar{\tau}}}
\newcommand{\mgio}{M_{\Gamma in,1}}
\newcommand{\Lt}{L_{\bar{\tau}}}
\newcommand{\mgo}{M_{\Gamma out}}
\newcommand{\Gs}{\Gamma^{\frac{1}{2}}}
\newcommand{\Gsi}{\Gamma^{-\frac{1}{2}}}

\newcommand{\Ges}{\Gamma_{\epsilon}^{\frac{1}{2}}}
\newcommand{\Gesi}{\Gamma_{\epsilon}^{-\frac{1}{2}}}
\newcommand{\mge}{M_{\Gamma_{\epsilon}}}

\newcommand{\mg}{M_{\Gamma}}
\newcommand{\nga}{\tilde{N}_{\Gamma}}
\newcommand{\ngo}{\tilde{N}_{\Gamma out}}
\newcommand{\ngi}{\tilde{N}_{\Gamma in}}
\newcommand{\vg}{\tilde{V}_{\Gamma}}
\newcommand{\ug}{{U}_{\Gamma}}
\newcommand{\Qg}{Q_{\Gamma}}

\newcommand{\agi}{A_{\Gamma in}}
\newcommand{\bgi}{B_{\Gamma in}}
\newcommand{\cgi}{C_{\Gamma in}}
\newcommand{\dgi}{D_{\Gamma in}}

\newcommand{\ai}{A_{in}}
\newcommand{\bi}{B_{in}}
\newcommand{\ci}{C_{in}}
\newcommand{\di}{D_{in}}

\newcommand{\nii}{\bar{N}_{i,in}}
\newcommand{\aii}{A_{i,in}}
\newcommand{\bii}{B_{i,in}}
\newcommand{\cii}{C_{i,in}}
\newcommand{\dii}{D_{i,in}}

\newcommand{\sys}[4]{{\left[\begin{BMAT}[2.2pt,1.3cm,0.5cm]{c3c}{c3c}  
#1 & #2 \\
#3 & #4
\end{BMAT}\right]}}

\newcommand{\syss}[4]{{\left[\begin{BMAT}[2.2pt,1.3cm,0.5cm]{c3c}{c3c}  
\scriptstyle #1 & \scriptstyle #2 \\
\scriptstyle #3 & \scriptstyle #4
\end{BMAT}\right]}}

\newcommand{\pow}[1]{\left\|#1\right\|_{\mathcal{P}}}
\newcommand{\norm}[2]{\left\|#1\right\|_{#2}}
\newcommand{\rh}[1]{\mathcal{RH}_{#1}}

\newcommand{\hlt}[1]{{#1}}

\title{\hlt{Mean-Square Stability and Stabilizability for LTI and Stochastic Systems Connected in Feedback}} 
\author{Junhui Li, Jieying Lu, and Weizhou Su
\thanks{Junhui Li, Jieying Lu, and Weizhou Su are with the School of Automation Science and Engineering, South China University of Technology, Guangzhou, 510641, P.R. China.
        {\tt\small l.junhui@mail.scut.edu.cn}, {\tt\small aujylu@scut.edu.cn}, {\tt\small wzhsu@scut.edu.cn}}%
}

\begin{document}

\maketitle

\begin{abstract}
In this paper, the feedback stabilization of a linear time-invariant (LTI) multiple-input multiple-output (MIMO) system cascaded by a linear stochastic system is studied in the mean-square sense.
Here, the linear stochastic system can model a class of correlated stochastic uncertainties such as channel uncertainties induced by packet loss and random transmission delays in networked systems.
By proposing a key parameter called coefficient of frequency variation to characterize the correlation of the stochastic uncertainties, we present a necessary and sufficient condition of the mean-square stability for this MIMO stochastic feedback system.
After then a necessary and sufficient condition for the mean-square stabilizability is provided, which reveals a fundamental limit imposed by the system's unstable poles, nonminimum-phase (NMP) zeros, relative degrees (input delays), and the coefficient of frequency variation of the stochastic uncertainties.
A numerical example is presented to illustrate the fundamental constraints in the mean-square stabilizability of MIMO networked systems with parallel communication channels.
\end{abstract}


\section{Introduction}\label{Sec:Introduction}

In the past few decades, a huge amount of research interests has been attracted on networked control systems (NCSs) owing to their great advantages in the field of engineering \cite{Hespanha2007,ZHANG2017Survey,Park2018}.
However, the use of communication networks usually introduces some stochastic phenomena such as random transmission delays, random data dropouts, and packet disorder into the control systems, which may degrade the system performance or jeopardize the system stability.
Thus, analyzing and mitigating the effects of those stochastic phenomena on NCSs have also received considerable attention.
Before then, the most related field may be stochastic systems with random state- or input-dependent multiplicative uncertainties, see, e.g., \cite{Willems1971Frequency,Lu2002MeanSquare,Qi2017Control}.
In \cite{Willems1971Frequency}, a necessary and sufficient condition of mean-square stability, named as mean-square small-gain theorem, is presented for single-input single output (SISO) linear systems with an input-dependent multiplicative uncertainty. This result is extended to MIMO linear stochastic systems by Lu and Skelton in \cite{Lu2002MeanSquare}.
In \cite{Qi2017Control}, fundamental conditions for mean-square stabilizability of generic linear systems with structured i.i.d. multiplicative uncertainties were developed.
The distinguish work in \cite{Elia2005Remote} bridged the gap between memoryless fading channels and input-dependent i.i.d multiplicative uncertainties, and showed that in a stochastic control framework, networked linear systems with memoryless fading channels can be modeled as linear systems with i.i.d. multiplicative uncertainties.
More results have been reported in a large amount of works, e.g., in \cite{Xiao2009Mean,GonzalezChen2019}.
Although the aforementioned stochastic control framework is successful in analysis and design of linear systems with i.i.d. multiplicative uncertainties or memoryless fading channels, it is no longer applicable in coping with channels with memory or correlated stochastic uncertainties.
In contrast to i.i.d. multiplicative uncertainties, much less results on the stability and stabilizability conditions of systems with correlated stochastic multiplicative uncertainties are available.

In general, random delay of data transmission in discrete-time NCSs is a typical example of those correlated stochastic uncertainties \cite{Zhang2020}.
Analogous to multipath channels \cite{Goldsmith2005Wireless}, uncertain channels with random delay were discussed for NSCs in
\cite{QuevedoJ_TAC2014,Su2017mean-square,Su2021meansquare,Li2021Stabilization,Dipankar2022Optimal,Xu2023,Chen2024}.
In \cite{Su2017mean-square} and \cite{Su2021meansquare}, an impulse response based model is proposed for communication channels with random transmission delays and packet dropouts. Based on this model, the mean-square stability and stabilizability criteria are presented for the networked systems.
\hlt{
The main contributions of this work are summarized as follows.
\begin{enumerate}[1)]
  \item A general impulse response based model is established for MIMO networked systems with correlated stochastic uncertainties such as random transmission delays and packet dropouts.
  \item The mean-square stability condition for MIMO systems with correlated stochastic uncertainties is derived by using the impulse response based model, which provides a generalized mean-square small-gain theorem.
  \item Mean-square stabilizability conditions, both necessary and sufficient, for a class of MP and NMP plants with correlated stochastic uncertainties are derived, which explicitly reveals the intrinsic effect of the plant's characteristic (i.e., unstable poles, input delay, and NMP zeros) and the coefficient of frequency variation of the correlated stochastic uncertainties on the closed-loop stabilizability.
\end{enumerate}}

This paper is organized as follows.
Section \ref{Sec:Problem_Formulation} introduces the impulse response based model for the correlated stochastic uncertainties under study.
Section \ref{Sec:MSS} presents the main results of the mean-square stability and stabilizability criteria to the system with the correlated stochastic uncertainties.
\ref{Sec:Numerical_example} applies to results to the NCS by a numerical example.
Section \ref{Sec:Conclusion} draws conclusions.

The notations used in this paper \hlt{are} mostly standard.
The notations $A^T, A^*, A^{-1}, A^{-*}, \rho(A)$ stand for, respectively, the transpose, complex conjugate transpose, inverse, inverse conjugate transpose, and spectral radius of appropriate matrix $A$.
$e_i$ is the $i$th column of the identity matrix with appropriate dimension.
$\mathbb{Z}$ is the set of integers.
${\E}$ and ${\Cov}$ are respectively the expectation and covariance operators of random variables.
For any transfer function $P(z)$, its state-space realization is denoted by
$P(z)=\sys{A}{B}{C}{D}$.
A real random process $\{u(k):k=0,1,\cdots\}$ is {quasi-stationary} if its autocorrelation matrix $R_u(\lambda):=\lim_{N\to \infty}\frac{1}{N}\sum_{k=0}^{N-1}\E\{u(k + \lambda)u^T(k)\}$ exists for all $\lambda \in \mathbb{Z}$.
If this is the case, the power spectral density (PSD) of $u(k)$ is defined as $S_u({e^{j\omega}}):=\sum_{\lambda=-\infty}^{\infty}R_u(\lambda)e^{-j\omega \lambda}$, and the average power of $u(k)$ is defined as $\pow{u} = \sqrt{R_u(0)}$.
With a slight abuse of terminology, $S_u(z)=\sum_{\lambda=-\infty}^{\infty} R_u(\lambda)z^{-\lambda}$ is also called the PSD of $u(k)$.

\section{System Modeling and Problem Formulation}
\label{Sec:Problem_Formulation}

\begin{figure}[!hbt]
\centering
\resizebox{0.5\linewidth}{!}{
\begin{tikzpicture}[auto, node distance=2cm, >=stealth', line width=0.75pt]
\node[sum] (sum) {};
\node[block, right of= sum, xshift=-0.3cm] (G) {$P$};
\node[block, below of= G, xshift=0cm] (K) {$K$};
\node[block, below of= sum, yshift=0.9cm, minimum width= 1.2cm, minimum height=0.6cm] (omega) {$\Delta$};
\draw[->] ($(sum.west) + (-0.7cm,0cm)$) -- node[near start](){$v$} (sum);
\draw[->] (sum) -- (G);
\draw[->] (G) -- node[near end]{$y$} ($(G)+(2.3cm,0cm)$);
\draw[->] (G) -- ++(1.6cm,0cm) |- (K);
\draw[->] (K) -| node[]{$u$} (omega);
\draw[->] (omega) -- node[near start]{$u_d$} node[pos=0.8]{$-$} (sum);
\end{tikzpicture}
}
\caption{The discrete-time stochastic feedback system}
\label{Fig:Stochastic_system}
\end{figure}
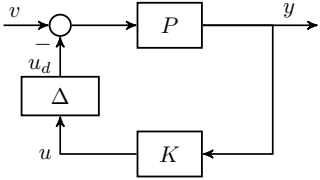

Consider a MIMO discrete-time stochastic feedback system shown in Fig. \ref{Fig:Stochastic_system}, where $P$ is an LTI system with $m$ inputs and relative degree greater than zero, and $K$ is an LTI controller with $m$ outputs.
$\Delta$ is a linear stochastic system with $m$ inputs and $m$ outputs, which may model the uncertainties with memory in the closed-loop such as an unreliable wireless communication channel over which the controller output $u$ is transmitted.
The signals $v,y$ and $u_d$ are the exogenous noise, the measurement of the plant $P$, and the output of $\Delta$, respectively.

Suppose that the stochastic system $\Delta$ consists of $m$ parallel subsystems $\Delta_1,\cdots,\Delta_m$, i.e.,
\begin{align}\label{Equ:Delta}
\Delta = \diag\{\Delta_1,\cdots,\Delta_m\}.
\end{align}
Denote by $u_i$ and $u_{di}$ respectively the input and output of the subsystem $\Delta_i$, $i=1,\cdots, m$.
Thus, the input and output of $\Delta$ are respectively written as
\begin{align*}
u=\begin{bmatrix} u_1 & \cdots & u_m   \end{bmatrix}^T~~
{\text{and}}~~ u_d=\begin{bmatrix} u_{d1} & \cdots & u_{dm}   \end{bmatrix}^T.
\end{align*}

In this work, our study focuses on the case that the stochastic system $\Delta$ is an unreliable communication channel whose unreliability is caused by random transmission delays.
Here, we adopt the impulse response based input-output model (see \cite{Su2017mean-square} and \cite{Su2021meansquare}) to describe each unreliable subchannel $\Delta_i$.
Denote by $\chi_{i,k}$ the time required to transmit the data $u_i(k)$ over $\Delta_i$.
Suppose the transmission time $\chi_{i,k}$ is a random variable taking values in the set $\mathcal{D}_i=\{0,1,\cdots, \bar{\chi}_i\}$.
Hence, at the instant $k$, the terminal of the subchannel $\Delta_i$ receives data belong to the set $\{u_i(k-j):j\in \mathcal{D}_i\}$ or no data is received, determined by the random variables $\{\chi_{i,k-j}:j\in \mathcal{D}_i\}$.
Since the terminal may receive more than one data during the sample interval, the linear combination of the received data is used as the output of $\Delta_i$, i.e.,
\begin{align}\label{Equ:Receiver_output}
u_{di}(k) = \sum_{j=0}^{\bar{\chi}_i}\alpha_{i, j}\delta(\chi_{i, k-j}-j)u_i(k-j),
\end{align}
where $\delta(\chi_{i, k-j}-j)$ indicates whether the data $u_i(k-j)$ is received at $k$ and $\alpha_{i, j}$ is a nonnegative weight assigned to the data $u_i(k-j)$ when it is received at $k$.

Now, we impose the following assumptions in the processes $\{\chi_{i,k}:k \in \Zset\},i=1,\cdots,m,$ of the stochastic system $\Delta$, which is aligned with the multipath wireless transmission experiments in \cite{Goldsmith2005Wireless} and the work \cite{Dipankar2022Optimal}.
\begin{assump}\label{Assum:vech_i}
It is assumed that 
\begin{enumerate}
\item for $i=1,\cdots, m$, each process $\{\chi_{i,k}\}$ is an independent and identically distributed (i.i.d.) process, and \label{Assum:i1}
\item the probability mass function (PMF) of $\chi_{i,k}$ is given by ${\Pr}\{\chi_{i,k}=j\}=p_{ij}, j\in \mathcal{D}_i$, with $\sum_{j\in \mathcal{D}_i}p_{ij} = 1$; \label{Assum:i2}
\item for any $i_1 \neq i_2$, the processes $\{\chi_{i_1,k}\}$ and $\{\chi_{i_2,k}\}$ are mutually independent. \label{Assum:i3}
\end{enumerate}
\end{assump}

Denote by $\{h_i(k,n): k \in \Zset \}$ the response of the subsystem $\Delta_i$ to the unit impulse input $u_i(k)=\delta(k-n)$.
Taking account of \eqref{Equ:Receiver_output}, we see that
$h_i(k,n)=\alpha_{k-n}\delta(\chi_{i,n}-(k-n)),k-n \in \mathcal{D}_i$ and
$h_i(k,n)\equiv 0,k-n \notin \mathcal{D}_i$.
That is, 
\begin{align}\label{FIR_IR}
&\left\{ h_i(k,n): k \in \Zset \right\}\nonumber\\
&=\big\{\cdots, 0,\mathop{\underline{\alpha_{i, 0}\delta(\chi_{i,n})}}\limits_{\mathop{\uparrow}  \limits_{k=n} }, \cdots, \mathop{\underline{\alpha_{i, \bar{\chi}_i}\delta(\chi_{i,n}-\bar{\chi}_i)}}\limits_{\mathop{\uparrow}  \limits_{k=n+\bar{\chi}_i} }, 0, \cdots\big\}.
\end{align}
Denote by $\vec{h}_i(n)$ the vector which consists of the nontrivial entries in the response sequence \eqref{FIR_IR}, i.e.,
\begin{align}
\vec{h}_i(n)=[h_i(n,n), \cdots, h_i(n+\bar{\chi}_i, n)]^T. \nonumber
\end{align}
It follows from (\ref{FIR_IR}) that the vector $\vec{h}_i(n)$ is related to $\chi_{i,n}$ only. Taking account of
Assumption \ref{Assum:vech_i}, we can see that the mean $\mu_i$ and covariance $R_i$ of $\vec{h}_i(n)$ are respectively a constant vector and a constant matrix, irrespective of $n$.
In other words, for all $n \in \Zset$,
\begin{align}\label{mean_covariance}
{\E}\{\vec{h}_i(n)\}=\mu_i~~ {\text{and}} ~~ {\Cov}\{\vec{h}_i(n)\}=R_i.
\end{align}

Denote the $j$th entry of $\mu_i$ by $\mu_{i,j-1}$ and the $(j_1,j_2)$th entry of $R_i$ by $r_{i,(j_1-1)(j_2-1)}$.
Let $\tilde{h}_i(k,k-j)=h_i(k,k-j)-\mu_{i,j}$. The input-output relation \eqref{Equ:Receiver_output} is rewritten as:
\begin{align}\label{Equ:Receiver_outputA}
u_{di}(k) = \sum_{j=0}^{\bar{\chi}_i}\mu_{i,j}u_i(k-j)+\sum_{j=0}^{\bar{\chi}_i}\tilde{h}_i(k,k-j)u_i(k-j).
\end{align}
It is shown by \eqref{Equ:Receiver_outputA} that the subsystem $\Delta_i$ is decomposed into a deterministic LTI part, denoted by $H_i$, and a correlated linear stochastic uncertainty, denoted by $\Omega_i$. The former is called the $i$th {\emph{nominal subsystem}} of $\Delta$, whose transfer function is
\begin{align}
H_i(z) = \sum \limits_{j=0}^{\bar{\chi}_i}\mu_{i,j} z^{-j}, \nonumber
\end{align}
and the latter, follows from \eqref{Equ:Receiver_outputA}, is characterized by the input-output relation:
\begin{align}
d_i(k) =  \sum_{j=0}^{\bar{\chi}_i} \tilde{h}_i({k,k-j}) u_i(k-j). \label{Equ:d_total}
\end{align}
Then the response of $\Omega_i$ to the unit impulse $\delta(k-n)$ is given by $\{ \cdots, 0,\tilde{h}_i(n,n), \cdots, \tilde{h}_i(n+\bar{\chi}_i, n), 0, \cdots\}$.
Similarly, denote by $\vec{\tilde{h}}_i(n)$ the vector constructed by the nontrivial entries in the response.
Indeed, it holds that $\vec{\tilde{h}}_i(n)=\vec{h}_i(n)-\mu_i$ and, by \eqref{mean_covariance},
\begin{align}\label{mean_covariance_Delta}
{\E}\{\vec{\tilde{h}}_i(n)\}=0~~ {\rm and} ~~ {\Cov}\{\vec{\tilde{h}}_i(n)\}=R_i.
\end{align}
Accordingly,
we define the energy spectral density function for the impulse response of $\Omega_i$ as below (see \cite{Su2021meansquare} for details):
\begin{align*} 
S_{\Omega_i}(z)=\sum_{\lambda=-\bar{\chi}_i}^{\bar{\chi}_i}r_i(\lambda)z^{-\lambda}
\end{align*}
where $r_i(\lambda) =\sum_{j = 0}^{\bar{\chi}_i-\lambda} r_{i,j(j+\lambda)}$ for $\lambda \in \mathcal{D}_i$ and $r_i(\lambda) = r_i(-\lambda)$ for $-\lambda \in \mathcal{D}_i$, and $r_i(\lambda) = 0$ for $|\lambda| \not\in \mathcal{D}_i$.

To reveal the input-output relation of $\Delta_i$ in the frequency domain, we make the following assumption which is standard in linear stochastic control area.
\begin{assump}\label{Assum:v}
The exogenous input sequence $\{v(k):k=0,1,\cdots\}$ is an i.i.d. white noise vector process with zero-mean and variance $\Sigma_v = \diag\{\sigma_{v_1}^2,\cdots,\sigma_{v_m}^2\}$, and is independent of the random processes $\{\chi_{i,k}\},i=1,\cdots,m$.
\end{assump}
Assumptions \ref{Assum:vech_i} and \ref{Assum:v} lead to two technical lemmas as below:
\begin{lma}[see \cite{Su2021meansquare}]\label{Lma:Independence_Singal}
For the plant $P$ with relative degree greater than zero in Fig. \ref{Fig:Stochastic_system}, $\chi_{i,k},i=1,\cdots, m$ are independent of $u(n),n<k$.
\end{lma}
\begin{lma}[see \cite{Su2021meansquare}]\label{Lma:InputOutputFeq}
For the correlated stochastic uncertainty $\Omega_i$, the PSD function of its output $d_i$ exists if and only if the averaged power of its input $u_i$ is a constant.
Moreover, the PSD of $d_i$, if exists, is given by
\begin{align}\label{Equ:S_d_u}
S_{d_i}(z)= S_{\Omega_i}(z) \|u_i\|_{\cal P}^2.
\end{align}
\end{lma}

\begin{rmk}
\label{Equ:Frequency_domain_description}
The frequency-domain model \eqref{Equ:S_d_u} of $\Omega_i$ covers several classical models, such as analog erasure and rice fading channels \cite{Elia2005Remote}, the channels subject to multiple random transmission delays \cite{Li2016Stabilization}, and classical i.i.d. multiplicative uncertainties \cite{Qi2017Control}.
It is significant that the model \eqref{Equ:S_d_u} can be generalized to the case when
{all entries on the right-hand side of \eqref{FIR_IR} are linear combinations of $\delta(\chi_{i,n}),\cdots,\delta(\chi_{i,n}-\bar{\chi}_i)$.}
This is a novel impulse response based description for a linear stochastic system, which characterizes the correlated stochastic uncertainty induced by randomly switching among a set of LTI systems.
{For this case, the input-output relation given by \eqref{Equ:S_d_u} also holds.}
\end{rmk}


\begin{figure}[!hbt]
\centering
\resizebox{0.5\linewidth}{!}{
\begin{tikzpicture}[auto, node distance=2cm, >=stealth', line width=0.7pt]
\node[block](P){$P$};
\node[sum, left of= P, xshift=0cm](sum1){};
\node[sum, left of= sum1, xshift=0.5cm](sum2){};
\node[block, below right of= P, xshift=0.2cm, yshift=0.6cm](K){$K$};
\node[block, below of= P, yshift=0.4cm, minimum width= 1.2cm, minimum height=0.6cm](H){$H$};
\node[block, below of= H, yshift=1cm, minimum width= 1.2cm, minimum height=0.6cm](D){$\Omega$};

\draw[->]($(sum2.west)+(-0.7cm,0cm)$) -- node[pos=0.2]{$v$} (sum2);
\draw[->](sum2) -- (sum1);
\draw[->](sum1) -- node[]{} (P);
\draw[->](P) -| (K);
\draw[->](P) --  node[pos=0.9]{$y$} ($(P.east)+(1.8cm,0cm)$);
\draw[->](K) |- (H);
\draw[->](H) -| node[pos=0.9]{$-$} (sum1); 
\draw[->](K) |-  node[near start]{$u$} (D);
\draw[->](D) -| node[near end, swap]{$d$} node[pos=0.95]{$-$} (sum2);

\end{tikzpicture}
}
\caption{LTI system with a stochastic uncertainty block}
\label{Fig:System_G_with_uncertainty}
\end{figure}
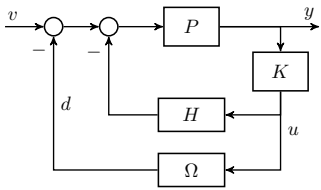

According to \eqref{Equ:Delta} and \eqref{Equ:Receiver_outputA}, the closed-loop system shown in Fig. \ref{Fig:Stochastic_system}
is reconstructed as that shown in Fig. \ref{Fig:System_G_with_uncertainty}, where $H$ is referred to as the \emph{nominal system} of $\Delta$ 
given by
\begin{align}
H(z) = \diag\{H_1(z),H_2(z),\cdots,H_m(z)\} \nonumber
\end{align}
and $\Omega$ is the stochastic uncertainty block of $\Delta$, i.e.,
\begin{align}
\Omega = \diag\{\Omega_1,\Omega_2,\cdots,\Omega_m\}. \nonumber
\end{align}

\begin{lma}\label{uncorrelated_di}
Suppose that the relative degree of the plant $P$ is greater than zero, and
Assumptions \ref{Assum:vech_i} and \ref{Assum:v} hold.
Then for any $k_1, k_2 \in \Zset$, $i_1,i_2=1,\cdots, m$
and $i_1 \neq i_2$, it holds that the outputs $d_{i_1}(k_1)$ and $d_{i_2}(k_2)$ of $\Omega_{i_1}$ and $\Omega_{i_2}$, respectively, are orthogonal, i.e., $\E\{d_{i_1}(k_1)d_{i_2}(k_2)\}=0$.  
\end{lma}
\begin{IEEEproof}
By \eqref{Equ:d_total}, it can be shown that $
\E\{d_{i_1}(k_1)d_{i_2}(k_2)\}
= \sum_{j_1=0}^{\bar{\chi}_{i_1}} \sum_{j_2=0}^{\bar{\chi}_{i_2}} \epsilon_{j_1,j_2}$,
where $ \epsilon_{j_1,j_2} = \E\{\tilde{h}_{i_1}(k_1,k_1-j_1) \tilde{h}_{i_2}(k_2,$ $k_2-j_2)
 u_{i_1}(k_1-j_1) u_{i_2}(k_2-j_2)\}$.
It follows from Assumption \ref{Assum:vech_i}.\ref{Assum:i3}, Lemma \ref{Lma:Independence_Singal}, and \eqref{mean_covariance_Delta} that, for $i_1 \ne i_2$ and $k_1 - j_1 \ge k_2 - j_2$,
\begin{align*}
\epsilon_{j_1,j_2}&=\E\{\tilde{h}_{i_1}({k_1,k_1\!-\!j_1})\} \\
&\hspace{1cm} \times \E\{\tilde{h}_{i_2}({k_2,k_2\!-\!j_2}) u_{i_1}(k_1\!-\!j_1) u_{i_2}(k_2\!-\!j_2)\}\\
&=0.
\end{align*}
The similar result can be shown for $k_1 - j_1 < k_2 - j_2$.
Hence, $\E\{d_{i_1}(k_1)d_{i_2}(k_2)\}=0$, which completes the proof.
\end{IEEEproof}

According to Assumptions \ref{Assum:vech_i}.\ref{Assum:i1}, \ref{Assum:vech_i}.\ref{Assum:i3} and Lemma \ref{uncorrelated_di}, it holds, for any $k_1, k_2$, that
\begin{align}
&{\E}\{d(k_1)d^T(k_2)\}\nonumber\\
&=\diag\left\{{\E}\{d_1(k_1)d_1^T(k_2)\}, \cdots,
{\E}\{d_m(k_1)d_m^T(k_2)\}\right\}. \nonumber
\end{align}
This leads to
$ S_d(z)=\diag\{ S_{d_1}(z),\cdots, S_{d_m}(z)\}. $
By applying (\ref{Equ:S_d_u}), $S_d(z)$ is rewritten as below:
\begin{align}\label{Equ:S_d_MIMO}
S_d(z)= S_{\Omega}(z) \diag\{ \|u_1\|_{\cal P}^2,\cdots,\|u_m\|_{\cal P}^2\}.
\end{align}
where $S_{\Omega}(z)$ is referred to as the energy spectral density function of $\Omega$ given by
\begin{align}\label{Equ:S_Omega_MIMO}
S_{\Omega}(z) = \diag\{S_{\Omega_1}(z),\cdots,S_{\Omega_m}(z)\}.  
\end{align}

{It follows from \cite{Zhou1995} that}
there would exist a MP polynomial matrix $\Phi(z) = \diag\{\Phi_1(z),\cdots,\Phi_m(z)\}$ in $z^{-1}$, with the order of $\Phi_i(z)$ being $\bar{\chi}_i$ determined by $S_{\Omega_i}(z)$,  such that
\begin{align}\label{Equ:Phi0}
S_{\Omega}(z)=\Phi(z)\Phi^\sim(z),
\end{align}
{where $\Phi^\sim(z)$ is a shorthand for $\Phi^T(z^{-1})$.
In fact, equality \eqref{Equ:Phi0}} is known as the spectral factorization of $S_{\Omega}$, see, e.g., \cite{Zhou1995}, and $\Phi(z)$ is called the spectral factor of $S_{\Omega}(z)$.

\begin{figure}[!hbt]
\centering
\resizebox{0.5\linewidth}{!}{
\begin{tikzpicture}[auto, node distance=2cm, >=stealth', line width=0.7pt]
\node[sum](sum){};
\node[block, right of= sum](P){$G$};
\node[block, below of= P, yshift=0.6cm](channel){$\Omega$};
%
\draw[->]($(sum)+(-1cm,0cm)$) -- node[near start]{$v$} (sum);
\draw[->](sum) -- node[near start]{$e$} (P);
\draw[->](P) -- node[near end]{$u$} ++(2.6cm,0cm);
\draw[->](P) -- ++(2cm,0cm) |- (channel);
\draw[->](channel) -| node[near end]{$d$} node[pos=0.95]{$-$} (sum);
\end{tikzpicture}
}
\caption{Equivalent interconnection with input-output signals}\label{Fig:G_Omega}
\end{figure}
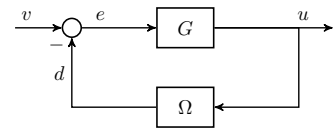

Define the nominal feedback system $G(z)$ to be the transfer function from $d$ to $u$ in Fig. \ref{Fig:System_G_with_uncertainty}, without considering $\Omega$, i.e.,
\begin{align*}
G(z) &= K(z)[I + P(z)H(z) K(z) ]^{-1}P(z). 
\end{align*}
In this case, Fig. \ref{Fig:System_G_with_uncertainty} is re-diagramed as Fig. \ref{Fig:G_Omega}. 

Let the set of all proper controllers internally stabilize $G(z)$ be $\mathcal{K}$.
Throughout this paper, we focus on the mean-square internal stability and stabilizability defined next.
\begin{defn}\label{Def:Internally_MS}
The stochastic feedback system illustrated in Fig. \ref{Fig:System_G_with_uncertainty} is said to be mean-square internally stable if the nominal feedback system $G(z)$ is internally stable and, in Fig. \ref{Fig:G_Omega}, for any white noise input $\{v(k)\}$ with bounded variance $\E\{v(k)v^T(k)\}<\infty$, the variances of $e(k)$ and $u(k)$ are also bounded, i.e., $\E\{e(k)e^T(k)\}<\infty$ and $\E\{u(k)u^T(k)\}<\infty$.
\end{defn}

\begin{defn}\label{Def:Mean_square_stabilizable}
The stochastic feedback system illustrated in Fig. \ref{Fig:System_G_with_uncertainty} is said to be mean-square stabilizable if there exists a controller $K \in \mathcal{K}$ such that the resulting system is mean-square internally stable.
\end{defn}

\section{Mean-Square Internal Stability and Stabilizability}
\label{Sec:MSS}

The coefficient of variation for a nonzero-mean random variable is defined as the ratio of its standard deviation to its mean \cite{Stewart2009Probability}.
Note that $\Phi(z)$ and $H(z)$ are the spectral factor and mean (nominal part) of the uncertainty $\Delta$, respectively.
We may analogously define the \emph{coefficient of frequency variation} of $\Delta$ to be
\begin{align}\label{Equ:W_total}
W(z) = H^{-1}(z)\Phi(z) = \diag\{W_1(z),\cdots,W_m(z)\},
\end{align}
where $W_i(z) = {H_i^{-1}(z)}{\Phi_i(z)}$ is referred to as the coefficient of frequency variation of $\Delta_i$.
By frequency variation, we mean that $W(e^{j\theta})$ would reflect the coefficient of variation of $\Delta$ at the frequency $\theta $ in the sense of spectral density.
This can be seen by setting the input of $\Delta$ as a white vector noise and then observing its output spectral in frequency $\theta$.
It is noteworthy that, since $\Phi_i(z)$ and $H_i(z)$ are $\bar{\chi}$th-order MP polynomials, $W_i(z)$ and $W(z)$ are invertible in $\rh{\infty}$.


\subsection{Mean-square internal stability}

Define the nominal complementary sensitivity function $T(z)$ corresponding to the nominal feedback system $G(z)$ as
\begin{align}\label{Equ:T_MIMO}
T(z) &= K(z)[I + P(z)H(z) K(z) ]^{-1}P(z)H(z). \nonumber 
\end{align}
\hlt{Denote $T_{i_1i_2}$ the element located in the $i_1$th row and the $i_2$th column of $T$, $i_1,i_2\in\{1,\cdots,m\}$}.
It is ready to state the criterion of the mean-square internal stability of the system.

\begin{thm}\label{Thm:MS_stable_condition}
The stochastic feedback system in Fig. \ref{Fig:Stochastic_system} is mean-square internally stable if and only if
\begin{equation}\label{Equ:MS_stable_condition}
\rho({\widehat{TW}})<1
\end{equation}
where
\begin{align*}
{\widehat{TW}} = {\begin{bmatrix}
{\left\| {{T_{11}}{W _1}} \right\|_2^2}& \cdots &{\left\| {{T_{1m}{W _m}}} \right\|_2^2}\\
 \vdots &{}& \vdots \\
{\left\| {{T_{m1}}{W _1}} \right\|_2^2}& \cdots &{\left\| {{T_{mm}{W_m}}} \right\|_2^2}
\end{bmatrix}}.
\end{align*}
\end{thm}

\begin{IEEEproof}
See Appendix \ref{Appendix:MS_stability}.
\end{IEEEproof}

\subsection{Mean-square stabilizability}

Let a right and left coprime factorization of the transfer function $PH$ be given by
\begin{equation*}
P(z)H(z) = N M^{-1} = \tilde{M}^{-1} \tilde{N},
\end{equation*}
where $N,M,\tilde{M},\tilde{N} \in \mathcal{RH}_\infty$ satisfy the double Bezout identity
\begin{align*}
\begin{bmatrix}
  \tilde{V} & \tilde{U} \\
  -\tilde{N} & \tilde{M}
\end{bmatrix}\begin{bmatrix}
               M & -U \\
               N & V
             \end{bmatrix} = \begin{bmatrix}
                               M & -U \\
                               N & V
                             \end{bmatrix}\begin{bmatrix}
  \tilde{V} & \tilde{U} \\
  -\tilde{N} & \tilde{M}
\end{bmatrix}
= I
\end{align*}
for some $U,V,\ut,\vt \in \mathcal{RH}_\infty$.
It is well-known that every stabilizing controller $K \in \mathcal{K}$ can be parameterized as \cite{Zhou1995}
\begin{align}
K &= (U+MQ)(V-NQ)^{-1} \nonumber\\
&= (\vt - Q \nt)^{-1}(\ut + Q\mt),~~Q \in \mathcal{RH}_\infty. \label{Equ:Youla_param_controller}
\end{align}
Substituting \eqref{Equ:Youla_param_controller} into the nominal complementary sensitivity function $T(z)$ yields
\begin{equation*}
T(z) = (U + MQ)\tilde{N},~~Q \in \mathcal{RH}_\infty.
\end{equation*}

In light of Theorem \ref{Thm:MS_stable_condition}, the following condition for mean-square stabilizability is immediate.
\begin{corollary}\label{Thm:Stabilizable_condition}
The stochastic feedback system in Fig. \ref{Fig:System_G_with_uncertainty} is mean-square stabilizable if and only if
\begin{equation}\label{Equ:Stabilizable_condition}
\rho_{\min} = \inf_{Q\in \rh{\infty}}\rho(J(Q))<1
\end{equation}
where
\begin{align*}
&J(Q)= \begin{bmatrix}\begin{smallmatrix}
{\| {[(U + MQ)\tilde{N}]_{11}}{W_{1}}\|_2^2}& \cdots &{\| {[(U + MQ)\tilde{N}]_{1m}{W_{m}}} \|_2^2}\\
 \vdots &{}& \vdots \\
{\| {[(U + MQ)\tilde{N}]_{m1}{W_{1}}} \|_2^2}& \cdots &{\| {[(U + MQ)\tilde{N}]_{mm}{W_{m}}} \|_2^2}
\end{smallmatrix}\end{bmatrix}.
\end{align*}
\end{corollary}

\Cref{Thm:Stabilizable_condition} makes it clear that to ascertain the mean-square stabilizability of the stochastic feedback system, one must solve a minimization problem for the spectral radius $\rho(J(Q))$ over $Q \in \rh{\infty}$.
The following lemma, which alternatively characterizes the spectral radius of positive matrices, serves as one of the technical base in the subsequent developments.

\begin{lma}[see \cite{Horn1986Matrix}]\label{Lma:Spectral_radius_lemma}
For any nonnegative matrix $J$,
\begin{align*}
\rho(J)=\inf_{\Gamma}\|\Gamma J \Gamma^{-1}\|_1 = \inf_{\Gamma}\|\Gamma J \Gamma^{-1}\|_\infty,
\end{align*}
where the infimum is taken over the set of positive diagonal matrices $\Gamma = \diag\{ \gamma_1^2, \cdots , \gamma_m^2 \}$.
\end{lma}

Conduct an inner-outer factorization $M = M_{in}M_{out}$ with $M_{in}=\sys{A_{in}}{B_{in}}{C_{in}}{D_{in}}$ and denote $\mg = \Gs M \Gsi$ for a given $\Gamma>0$.
Notice that $\mg =(\Gs \mi) (\mo \Gsi)$, the rational functions $\mg$ 
can be factorized as $\mg = M_{\Gamma in}M_{\Gamma out}$ with $\mgi$ being the inner of $\Gs M_{in}$ that \cite{Lin1996Inner}
\begin{align}
&\mgi = \sys{\agi}{\bgi}{\cgi}{\dgi} \label{Equ:M_Gamma_in} \\
&=
\sys{ \ai + \bi F }{ \bi(\di^*\Gamma \di \!+\!  \bi^*X\bi)^{-\frac{1}{2}} }{ \Gs \ci + \Gs \di F }{\Gs \di (\di^*\Gamma \di \!+\! \bi^*X\bi)^{-\frac{1}{2}}} \nonumber
\end{align}
with $X>0$ being the solution to the discrete algebraic Riccati equation (DARE)
\begin{align}
X &= \ai^* X\ai  + \ci^*\Gamma\ci - (\ai^*X\bi + \ci^*\Gamma\di) \nonumber \\
&\hspace{0.5cm}\times(\di^*\Gamma \di + \bi^*X\bi)^{-1}(\bi^*X\ai+\di^*\Gamma\ci) \label{Equ:X}
\end{align}
and $ F \!= \!-(\di^*\Gamma \di \!+\! \bi^*X\bi)^{-1}(\bi^*X\ai \!+\! \di^*\Gamma\ci)$.
Moreover, by \cite{Zhou1995},
\begin{align}\label{Equ:Mgii_realization}
\mgi^{-1} = \sys{\hat{A}}{\hat{B}}{\hat{C}}{\hat{D}} = \sys{A_{in} - \bi \di^{-1}\ci}{-\bgi \dgi^{-1}}{\dgi^{-1}\cgi}{\dgi^{-1}}.
\end{align}
Denote the Laurent expansion of the $i$th column of $\mgi^{-1}$ 
at infinity by
\begin{align}\label{Equ:Mgi_Laurent}
\mgi^{-1} e_i &= \sum_{j=0}^{\infty} \phi_{ji} z^{-j}
\end{align}
and define $f_{\tau,i} = \sum_{j=0}^{\tau-1} \phi_{ji} z^{-j}$ for a given integer $\tau\ge 1$.
Denote by $\mathbf{\Pi}_1$ the projection on the ${\mathcal{H}_2^\perp}$ space.
The following result reveals the interaction of $\mgi^{-1} e_i$ with the $\mathcal{H}_2$ space.

\begin{lma}\label{Lma:M-f}
For any scalar system $\hat{G}(z) \in \mathcal{H}_2$, it holds that
\begin{align}
Z_{1i} :=&~ \mathbf{\Pi}_1\{(f_{\tau,i}-\mgi^{-1}e_i)z^{\tau} \hat{G}(z)\} \nonumber\\
=&~ \hat{C} (zI-\hat{A})^{-1} \hat{A}^{\tau-1} \hat{G}(\hat{A}) \hat{B} e_i z \label{Equ:Z1i_proj}
\end{align}
and $\norm{Z_{1i}}{2}^2= e_i^*\hat{B}^*\hat{G}^*(\hat{A})\hat{A}^{*(\tau-1)}X \hat{A}^{\tau-1}\hat{G}(\hat{A})\hat{B}e_i$,
where $X$ is the solution to the DARE \eqref{Equ:X}.
\end{lma}

\begin{IEEEproof}
See Appendix \ref{Appendix:M-f}.
\end{IEEEproof}


\subsubsection{Minimum phase plants with input delays}

Consider a plant with input delays, which can be described by delays in the transfer matrix $\nt(z)$:
\begin{align}
  \nt(z) &= \nto(z)\diag\{ z^{-\tau_1},\cdots, z^{-\tau_m} \} \label{Equ:Input_delays}
\end{align}
where $\tau_i \ge 1,i=1,\cdots,m$.
Here, we assume that $\nto^{-1} \in \rh{\infty}$.
It should be noted that the structure of input delays given by \eqref{Equ:Input_delays} covers the case that $P(z)$ is of relative degree $\tau_0$, i.e., $\lim_{|z|\to \infty}\nt z^{\tau_0}$ is invertible.
For convenient, we shall define $\Lt = \diag\{z^{\tau_1},\cdots,z^{\tau_m}\}$.
A computational efficient solution to the mean-square stabilizability problem is ready to be presented.

\begin{thm}\label{Thm:Stabilizability_condition_delay}
Suppose that $P(z)$ is minimum phase and has input delays given by \eqref{Equ:Input_delays}.
Let $M = \mi \mo$ be an inner-outer factorization of $M$, with $\mi = \sys{A_{in}}{B_{in}}{C_{in}}{D_{in}}$ being an inner of $M$.
Then
\begin{align}
  \rho_{\min} &= \inf_{\Gamma}\Big\{ \mu: e_i^*\di^{-*}\bi^* W_i(\hat{A}^*){\hat{A}^{*^{\tau_i-1}}} X {\hat{A}}^{\tau_i-1} \nonumber \\
  &\hspace{0.2cm}  \times  W_i(\hat{A})\bi \di^{-1}e_i \le \mu e_i^*\Gamma e_i , i=1,\cdots,m \Big\} \label{Equ:rho_min0}
\end{align}
where $X>0$ is the solution to the DARE \eqref{Equ:X}.
Furthermore,
\begin{align}
&\rho_{\min} = \inf_{\gamma_1,\cdots,\gamma_m}\Big\{\mu: \sum_{i=1}^{m} \gamma_i X_i  > \frac{\gamma_i }{\mu} {\hat{A}}^{\tau_i-1}W_i(\hat{A}) \bi \di^{-1} e_i \nonumber\\
&\hspace{0.0cm}\times e_i^* \di^{-*}\bi^* W_i(\hat{A}^*){(\hat{A}^*)}^{\tau_i-1},\gamma_i>0,i=1,\cdots,m \Big\} \label{Equ:rho_min_2}
\end{align}
where $X_i \ge 0,i=1,\cdots,m$ is the solution to the Lyapunov equation
\begin{align}\label{Equ:X_i}
 X_i = \hat{A} X_i \hat{A}^{*} - \bi \di^{-1} e_i e_i^* \di^{-*} \bi^*.
\end{align}
The stochastic feedback system is mean-square stabilizable is and only if $\rho_{\min}<1$.
\end{thm}

\begin{IEEEproof}
By Corollary \ref{Thm:Stabilizable_condition} and Lemma \ref{Lma:Spectral_radius_lemma}, together with the Bezout identity that $M\vt + U\nt = I$, we have
\begin{align}
\rho(J(Q)) 
&= \inf_{\Gamma}\max_i \|\Gamma^{\frac{1}{2}}(I-M\vt + MQ\nt)\Gsi W e_i\|_2^2 \nonumber
\end{align}
Let $\nga =\Gs \nt \Gsi, \vg = \Gs \vt \Gsi,\Qg = \Gs Q \Gsi$ and $J_{\Gamma i}(\Qg) = \| (I-\mg \vg \!+\! \mg \Qg \nga) W e_i \|_2^2$.
Then
\begin{align}
\rho_{\min} 
&= \inf_{\Gamma}\max_i \inf_{\Qg \in \rh{\infty}}  J_{\Gamma i}(\Qg). \label{Equ:rho_min_inf}
\end{align}
It follows from $\mg = M_{\Gamma in}M_{\Gamma out}$ that
\begin{align}
J_{\Gamma i}(\Qg)= \| (\mgi^{-1}-\mgo \vg + \mgo \Qg \nga) W e_i \|_2^2 &\nonumber \\
= \| [(\mgi^{-1}-\mgo \vg)\Lt + \mgo \Qg \ngo ] W e_i \|_2^2 &\nonumber
\end{align}
where $\ngo = \nga \Lt = \Gs \nto \Gsi$ is invertible in $\rh{\infty}$.
Note that the Bezout identity $M\vt + U\nt = I$ implies that
\begin{align}\label{Equ:MMV}
\mgi^{-1}(I-\ug\nga ) e_i = \mgo \vg e_i
\end{align}
where $\ug = \Gs U\Gsi$.
Since $\nga e_i = \ngo e_i z^{-\tau_i}$ and $\mgi^{-1}$ and $\mgo \vg$ have relative degree zero, 
$\mgo \vg e_i$ has the Laurent expansion at infinity of the form
\begin{align*}
\mgo \vg e_i &= 
f_{\tau_i,i} + \sum_{j=\tau_i}^{\infty} \varphi_{ji} z^{-j} 
\end{align*}
\hlt{where $f_{\tau_i,i}$ is determined by the Laurent expansion of $\mgi^{-1}e_i$ in \eqref{Equ:Mgi_Laurent}.}
Therefore, by letting ${M}_{\Gamma in,\bar{\tau}}^{-1} = \begin{bmatrix}
                                   f_{\tau_1,1} & \cdots & f_{\tau_m,m}
                                  \end{bmatrix}$,
one can see that $(\mgit^{-1}-\mgo \vg)\Lt \in \mathcal{RH}_2$.
Accordingly, 
\begin{align}
J_{\Gamma i}(\Qg)&=\big\|(\mgi^{-1}-\mgit^{-1}) \Lt W e_i \nonumber \\
&\hspace{-1cm}+ \big[(\mgit^{-1} -  \mgo\vg)\Lt+ \mgo \Qg \ngo \big] W  e_i \big\|_2^2. \nonumber
\end{align}
Next, by conducting a stable-unstable decomposition
\begin{align}\label{Equ:M_Gamma_decompose}
(\mgi^{-1}-\mgit^{-1}) \Lt W = Z_1 + Z_2
\end{align}
where $Z_1 \in \mathcal{H}_2^{\perp}$ and $Z_2\in \mathcal{H}_2$, we obtain
\begin{align}
J_{\Gamma i}(\Qg)
&= \| Z_1 e_i\|_2^2 + \big\| (\bar{L}_{\Gamma} + \mgo \Qg \ngo W) e_i \big\|_2^2  \nonumber 
\end{align}
where $\bar{L}_{\Gamma} = Z_2 +  (\mgit^{-1} -  \mgo\vg)\Lt W \in \rh{\infty}$.
Since both $\mgo,\ngo$ and $W$ are invertible in $\rh{\infty}$, we have 
\begin{align}\label{Equ:inf_Q_Z1}
\inf\limits_{\Qg \in \rh{\infty}}J_{\Gamma i}(\Qg) =\left\| Z_1 e_i \right\|_2^2,
\end{align}
where $\Qg = -\mgo^{-1} \bar{L}_{\Gamma} W^{-1}\ngo^{-1}\in \rh{\infty}$ achieves the infimum.
\hlt{It should be noted that $\Qg$ is irrelevant to $\{e_i:i=1,\cdots,m\}$ so that $J_{\Gamma i}(\Qg),i=1,\cdots,m,$ are simultaneously minimized.}
On the other hand, it follows from \eqref{Equ:M_Gamma_decompose} that $Z_{1}e_i = \mathbf{\Pi}_1\{(\mgi^{-1}e_i - f_{\tau_i,i})z^{\tau} W_i\}$.
Then applying \Cref{Lma:M-f} with $\hat{G} = W_i$ yields
\begin{align}
&\|Z_1e_i\|_2^2 = e_i^*\hat{B}^* W_i(\hat{A}^*){\hat{A}^{*^{\tau_i-1} }} X {\hat{A}}^{\tau_i-1}W_i(\hat{A})\hat{B}e_i.   \label{Equ:Z_1_H2_Norm}
\end{align}
Combining \eqref{Equ:rho_min_inf}, \eqref{Equ:inf_Q_Z1}, \eqref{Equ:Z_1_H2_Norm} and  $\hat{B} \!=\! \bgi \dgi^{-1} \!=\! \bi \di^{-1} \Gamma^{-\frac{1}{2}}$ establishes \eqref{Equ:rho_min0}.
The proof of the rest, namely \eqref{Equ:rho_min_2}-\eqref{Equ:X_i}, can refer to \cite{Qi2017Control}.
\end{IEEEproof}

\begin{rmk}
It is important to note that the stabilizability conditions in \Cref{Thm:Stabilizability_condition_delay} is a GEVP problem,
which can be efficiently solved using LMI optimization techniques \cite{Boyd1994Linear}.
\Cref{Thm:Stabilizability_condition_delay} also shows that the difficulty of stabilization increases exponentially with the input delays.
It is worth noting that $W_i(\hat{A})$, the interaction between the coefficient of frequency variation and the unstable poles of the plant, plays an important role in the stabilizability criterion.
If the stochastic system $\Delta$ is reduced to be an i.i.d. diagonally-structured multiplicative noise, it turns out that $W_i(\hat{A}) = \frac{\sigma_i}{\mu_i}$ such that \eqref{Equ:rho_min0} becomes \cite[Equality (17)]{Qi2017Control}.
In addition to the locations of the unstable poles, the realizations of the inner factor $\mi$ also depend on their directions, and as such, so do the stabilizability conditions.
In particular, when $\bi \di^{-1}e_i = 0$, the Lyapunov equation \eqref{Equ:X_i} yields a trivial solution $X_i = 0$, and hence
the $i$th inequality in \eqref{Equ:rho_min_2} is rendered moot; in other words, the coefficient of frequency variation $W_i(z)$ can be arbitrary and the stochastic subsystem $\Delta_i$ has no effect on the closed-loop stabilizability.
\end{rmk}

\subsubsection{Nonminimum phase plants}

Now we consider the case which $P$ has NMP zeros $s_1,\cdots,s_m$.
Each of them is associated with a column of $P$, i.e.,
\begin{align}\label{Equ:P_nonminimum}
  P &= P_0 \diag\Big\{\frac{z-s_1}{z},\cdots,\frac{z-s_m}{z}\Big\}
\end{align}
where $P_0$ is \hlt{of minimum phase} and with relative degree one.
\hlt{Note that the decoupling of NMP input zeros usually occurs in multi-path transmission, see, e.g., \cite{Lu2019Mean-square}.}
It should be noted that the results for the framework of \eqref{Equ:P_nonminimum} can be extended to the case that $ P = P_0 \diag\{ z^{-\tau_1}P_1,\cdots,z^{-\tau_m}P_m \}$,
where the scalar transfer functions $P_i,i=1,\cdots,m$ have more than one NMP zeros and relative degree zero, $\tau_i,i=1,\cdots,m$ are positive integers. 

\begin{thm}\label{Thm:Rho_min_nonminimum}
Suppose the plant $P$ is given by \eqref{Equ:P_nonminimum}.
Then
\begin{align}
\rho_{\min} &= \inf_{\Gamma}\Big\{ \mu: e_i^*\di^{-*}\bi^* W_i(\hat{A}^*) \nii(\hat{A}^*) X  \nii(\hat{A}) \nonumber \\
&\hspace{0.5cm}  \times W_i(\hat{A}) \bi \di^{-1}e_i \le \mu e_i^*\Gamma e_i , i=1,\cdots,m \Big\} \label{Equ:rho_min_nonminimum}
\end{align}
where $\nii(\hat{A}) = (I - s_i^*\hat{A}) (\hat{A} - s_i I)^{-1}$
and $X>0$ is the solution to the DARE \eqref{Equ:X}.
Furthermore,
\begin{align*}
&\rho_{\min} = \inf_{\gamma_1,\cdots,\gamma_m}\Big\{\mu: \sum_{i=1}^{m} \gamma_i X_i  > \frac{\gamma_i }{\mu} \nii(\hat{A}) W_i(\hat{A}) \bi \di^{-1} \\
&\hspace{0.5cm}\times  e_i e_i^* \di^{-*}\bi^* W_i(\hat{A}^*)\nii(\hat{A}^*),\gamma_i>0,i=1,\cdots,m \Big\}
\end{align*}
where $X_i \ge 0,i=1,\cdots,m$ is the solution to the Lyapunov equation \eqref{Equ:X_i}.
The stochastic feedback system is mean-square stabilizable if and only if $\rho_{\min}<1$.
\end{thm}

\begin{IEEEproof}
Mimicking the proof of Theorem \ref{Thm:Stabilizability_condition_delay} shows that
\begin{align}
&J_{\Gamma i}(\Qg) = \| Z_1 e_i \|_2^2  \nonumber \\
&\hspace{0.5cm}+ \| [Z_2 + (\mgio^{-1} -\mgo \vg + \mgo \Qg \nga)Wz] e_i \|_2^2 \nonumber
\end{align}
where $Z_1$ and $Z_2 $ are, respectively, the $\mathcal{H}_2^{\perp}$ and $\mathcal{H}_2$ parts of $(\mgi^{-1}-\mgio^{-1}) W z$ with $\mgio^{-1} = \begin{bmatrix}
                f_{1,1} & \cdots & f_{1,m}
              \end{bmatrix} = \mgi^{-1}(\infty)$.
We then conduct the co-inner-outer factorization $\nga = \ngo \ngi$ where $\ngi$ is co-inner.
Clearly, $z\ngo$ is invertible in $\rh{\infty}$ and $\ngi e_i = \frac{z - s_i}{s_i^*z -1} e_i$ since \eqref{Equ:P_nonminimum}.
Thus, 
\begin{align*}
&\|[Z_2 + (\mgio^{-1}  -\mgo \vg + \mgo \Qg \nga)Wz] e_i\|_2^2 \\
&= \Big\|[Z_2 + (\mgio^{-1}  -\mgo \vg)Wz] \frac{s_i^*z -1}{z - s_i} e_i \\
&\hspace{3cm}+ \mgo \Qg \ngo W z e_i \Big\|_2^2 \\
&=  \Big\|[Z_2(s_i) + (\mgio^{-1}  -\mgo(s_i) \vg(s_i))W(s_i)s_i]\\
&\hspace{0.8cm}\times \frac{(|s_i|^2 -1)z}{s_i(z - s_i)} e_i \Big\|_2^2 + \| L_i + \mgo \Qg \ngo Wz e_i \|_2^2
\end{align*}
where $L_i$ is the $\mathcal{H}_2$ part of the partial fraction decomposition of $[Z_2 + (\mgio^{-1}  -\mgo \vg)Wz] \frac{s_i^*z -1}{z - s_i} e_i$.
Let $L = \begin{bmatrix}
           L_1 & \cdots & L_m
         \end{bmatrix}$.
Since $\mgo,z\ngo$ and $W$ are invertible in $\rh{\infty}$, $\Qg = -\mgo^{-1} L W^{-1} (z\ngo)^{-1} \in \rh{\infty}$ simultaneously minimizes $J_{\Gamma i}(\Qg),i=1,\cdots,m$ such that
\begin{align*}
&\inf_{\Qg \in \rh{\infty}} J_{\Gamma i}(\Qg)= \| Z_1 e_i \|_2^2 + \frac{|s_i|^2 -1}{|s_i|^2}\|[Z_2(s_i) \\
&\hspace{1cm}+ (\mgio^{-1}  -\mgo(s_i) \vg(s_i))W(s_i)s_i] e_i \|_2^2.
\end{align*}
By the Bezout identity $M\vt + U\nt = I$ and the fact that $s_i$ is a zero of $\nt$ and $\nga$, corresponding to the direction $e_i$, we have $\mgo(s_i)\vg(s_i)e_i = \mgi^{-1}(s_i) e_i$.
Together with $(\mgi^{-1}(s_i)-\mgio^{-1} )W(s_i)s_i = Z_1(s_i) + Z_2(s_i)$, it holds that
\begin{align*}
&Z_2(s_i)e_i + [\mgio^{-1}  -\mgo(s_i) \vg(s_i)]W(s_i)s_i e_i \\
&= - Z_1(s_i)e_i.
\end{align*}
As a result,
\begin{align}\label{Equ:Ji_Q_NMP}
&\inf_{\Qg \in \rh{\infty}} J_{\Gamma i}(\Qg)= \| Z_1 e_i \|_2^2 + \frac{|s_i|^2 -1}{|s_i|^2}\| Z_1(s_i)e_i \|_2^2.
\end{align}

It is observed from \eqref{Equ:Z1i_proj} that $Z_1(s_i)e_i = -\hat{C} W_i(\hat{A})(s_i I - \hat{A})^{-1}\hat{B}e_i s_i$ so that
\begin{align*}
\|Z_1(s_i)e_i\|_2^2 &={|s_i|^2}  e_i^*\hat{B}^* W_i(\hat{A}^*) (s_i I - \hat{A})^{-*} \hat{C}^* \\
&\hspace{1cm}\times \hat{C} (s_i I - \hat{A})^{-1}  W_i(\hat{A})\hat{B} e_i. 
\end{align*}
Combining $\| Z_1 e_i \|_2^2$ given by \eqref{Equ:Z_1_H2_Norm} and the fact that $\hat{C}^*\hat{C} = \hat{A}^*X\hat{A}-X$ follows
\begin{align*}
&\inf_{\Qg \in \rh{\infty}} J_{\Gamma i}(\Qg) = \| Z_1 e_i \|_2^2 + \frac{|s_i|^2 -1}{|s_i|^2}\| Z_1(s_i)e_i \|_2^2 \\
&= e_i^*\hat{B}^* W_i(\hat{A}^*)(s_i I - \hat{A})^{-*} \\
&\hspace{0.5cm}\times [(s_i I - \hat{A})^{*}X(s_i I - \hat{A}) + ({|s_i|^2 -1})(\hat{A}^*X\hat{A}-X)] \\
&\hspace{0.5cm}\times (s_i I - \hat{A})^{-1}  W_i(\hat{A})\hat{B} e_i  \\
&= e_i^*\hat{B}^* W_i(\hat{A}^*)(s_i^* I - \hat{A}^*)^{-1}  (s_i \hat{A}^* - I)X\\
&\hspace{0.5cm}\times (s_i^* \hat{A} - I) (s_i I - \hat{A})^{-1}  W_i(\hat{A})\hat{B} e_i.
\end{align*}
By $\hat{B} \!=\! \bgi \dgi^{-1} \!=\! \bi \di^{-1} \Gamma^{-\frac{1}{2}}$, the proof is completed.
\end{IEEEproof}

\begin{rmk}
Similar to \Cref{Thm:Stabilizability_condition_delay}, \Cref{Thm:Rho_min_nonminimum} provides the necessary and sufficient condition for mean-square stabilizability of NMP MIMO plant in the form \eqref{Equ:P_nonminimum} with the stochastic system $\Delta$.
It reveals the effect of interaction between the NMP zeros and the unstable poles $\hat{A}$ on the stabilizability of the system, under the condition that the NMP zeros are input-decoupled.
Since $\nii(\hat{A}) = (I - s_i^*\hat{A}) (\hat{A} - s_i I)^{-1}$, when the NMP zero $s_i$ approaches one or more unstable poles of the plant, the matrix $(\hat{A} - s_i I)$ would become singular such that $\nii(\hat{A})$ is a ill-conditioned matrix and $\mu$ in \eqref{Equ:rho_min_nonminimum} is probably larger than 1.
In this case, the system is unstabilizable.
\end{rmk}

Note that some quasiconvex optimizations in \Cref{Thm:Stabilizability_condition_delay} and \Cref{Thm:Rho_min_nonminimum} should be solved for the mean-square stabilizability criteria.
To mitigate the essential complexity, a sufficient stabilizability condition would be presented, which does not require solving optimization problems.
To this end, we shall introduce a lemma on the upper triangular coprime factorization of a given plant $G_p$.

\begin{lma}[see \cite{Lu2019Mean-square}]\label{Lma:Upper_triangular_coprime_fact}
For a given plant $G_p$, there exists a coprime factorization pair $(N,M)$ such that $G_p = NM^{-1}$, and the transfer function matrix $M$ is upper triangular, i.e.,
\begin{align*}
  M &= \begin{bmatrix}\begin{smallmatrix}
         M_{11} &  M_{12} & \cdots &  M_{1m} \\
         0 &  M_{22} & \cdots &  M_{2m} \\
         \vdots & \vdots  & \ddots & \vdots \\
         0 & 0 & \cdots &  M_{mm}
       \end{smallmatrix}\end{bmatrix}
\end{align*}
with the diagonal elements $M_{ii},i=1,\cdots,m$ being inner functions, namely $M_{ii}^\sim M_{ii} = 1$.
\end{lma}

Now we conduct the upper triangular coprime factorization $PH=NM^{-1}$ and let $M_{ii} = \syss{\aii}{\bii}{\cii}{\dii}$ be a balanced realization of $M_{ii}$.

\begin{corollary}\label{Cor:Stabilizability_nonminimum_triangular}
Suppose the plant $P$ is given by \eqref{Equ:P_nonminimum}.
Then the stochastic feedback system is mean-square stabilizable if
\begin{align*}
 &\nii(\aii^{-*}) \bii \dii^{-1} \dii^{-*}\bii^*  \nii(\aii^{-1}) \\
 &\hspace{4cm}<  W_i^{-1}(\aii^{-*}) W_i^{-*}(\aii^{-*})
\end{align*}
hold for $i=1,\cdots,m$.
\end{corollary}

\begin{IEEEproof}
See Appendix \ref{Appendix:Stabilizability_triangular}.
\end{IEEEproof}

%
%
%

\section{Numerical Example}
\label{Sec:Numerical_example}

In this section, the proposed results would be verified by their application to NCSs with random transmission delays.
Consider the double-input and double-output plant
\begin{equation*}
P(z) = \begin{bmatrix}
         \frac{z-s_1}{z-\lambda} &  -\frac{z-s_2}{z-\lambda} \\
          0 &  \frac{z-s_2}{z}
       \end{bmatrix}\begin{bmatrix}
        z^{-\tau_1} &  0 \\
          0 &  z^{-\tau_2}
       \end{bmatrix},~~\lvert\lambda\rvert >1,
\end{equation*}
which has \hlt{one unstable pole at $z = \lambda$ with the input direction $\eta = \frac{1}{\sqrt{(\lambda-s_1)^2 + (\lambda-s_2)^2}}\begin{bmatrix}
                                                                                              \lambda-s_1 & -\lambda+s_2
                                                                                            \end{bmatrix}^*$} such that
\begin{align*}
\mi= \sys{A_{in}}{B_{in}}{C_{in}}{D_{in}}=
\syss{ {\lambda^{-*}}}{{ {\lambda^{-*}}{\sqrt{|\lambda|^2\!-\!1}}}\eta^*}{{ {\lambda^{-*}}{\sqrt{|\lambda |^2\!-\!1}}}\eta}{I\!-\! (1\!+\!{\lambda^{-*}}) \eta\eta^*}.
\end{align*}
The NMP zeros of the plant are $z=s_1$ and $z = s_2$ if $|s_1|>1$ and $|s_2|>1$, and the input delays are $\tau_1$ and $\tau_2$.

\subsection{Example 1}

\begin{figure}[!htb]
  \centering
  \includegraphics[width=7.5cm]{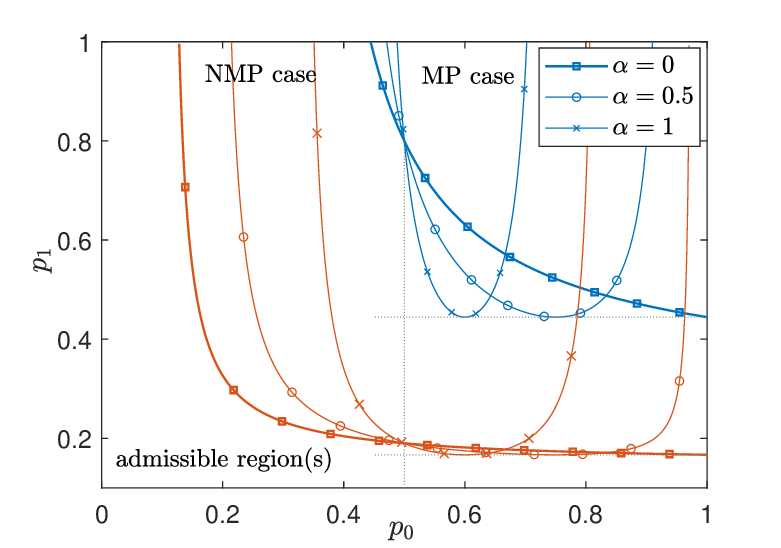}
  \caption{Mean-square stabilizable regions}\label{Fig:p0p1}
\end{figure}

Suppose now two individual communication channels are placed in the path from the controller to the plant.
The first channel is a one-step random transmission delay channel with linear receiving strategy, see \eqref{Equ:Receiver_output}.
The delay parameters are assumed to be $\Pr\{\tau_k = 0\} =  p_{0},\Pr\{\tau_k = 1\} = 1- p_{0}$ and the received signal weights/fadings $\alpha_{0} = 1,\alpha_{1}=\alpha$.
The second channel is a packet dropout channel with loss rate $p_1$.
Under this assumption,
\begin{align*}
W_1 = \sqrt{\frac{p_0}{1-p_0}}\frac{1-\alpha z^{-1}}{1-\alpha \frac{p_0}{1-p_0} z^{-1}} ,~~W_2 = \sqrt{\frac{p_1}{1-p_1}}.
\end{align*}
Further suppose that $\tau_1 = \tau_2 = 1$. 
It follows from Theorem \ref{Thm:Rho_min_nonminimum} that the system is mean-square stabilizable if and only if
\begin{align}\label{Equ:Lambda_p0_p1}
{| \lambda |^2 - 1} < n_{s_1} \mathbf{R}_{p_0}\bigg( \frac{\lambda - \alpha \mathbf{R}_{p_0}^{-1}}{\lambda - \alpha} \bigg)^2 + n_{s_2} \mathbf{R}_{p_1},
\end{align}
where $\mathbf{R}_{p_i} = {\frac{1-p_i}{p_i}}$ and $n_{s_i} = \frac{\lambda-s_i}{1-s_i^* \lambda}$ for $i=0,1$.
The admissible region(s) of the delay and packet dropout probabilities are numerically illustrated in Fig. \ref{Fig:p0p1}, for the case $\lambda=1.5$ and $s_1=3,s_2=4$ and the case without NMP zeros.
It is observed that the admissible region of the probabilities for mean-square stabilizability is strongly compressed by the NMP zeros of the plant, for a given $\alpha$.
Interestingly, there is a critical point in the boundary of each admissible region with $\alpha \ne 0$.
If the delay probability of the first channel is smaller than its critical value, an increase in it would decrease the admissible packet dropout probability of the second channel. 
Differently, if the delay probability is large enough, the admissible packet dropout probability would also be enlarged.
The reason is that adopting the linear receiving strategy changes the monotonicity of the effect of the delay probability on the stabilizability, see from comparing the first term on the right-hand side of the inequality \eqref{Equ:Lambda_p0_p1} with the second term. 
In addition, since $\mathbf{R}_{0.5}=1$, the boundaries of the admissible regions have an intersection point $(\bar{p}_0,\bar{p}_1)=(\frac{1}{2}, \frac{n_{s_2}}{|\lambda|^2-1-n_{s_1}+n_{s_2}})$, which is determined by the plant's unstable pole and the NMP zeros.

\subsection{Example 2}

\begin{figure}[!htb]
  \centering
  \includegraphics[width=7.5cm]{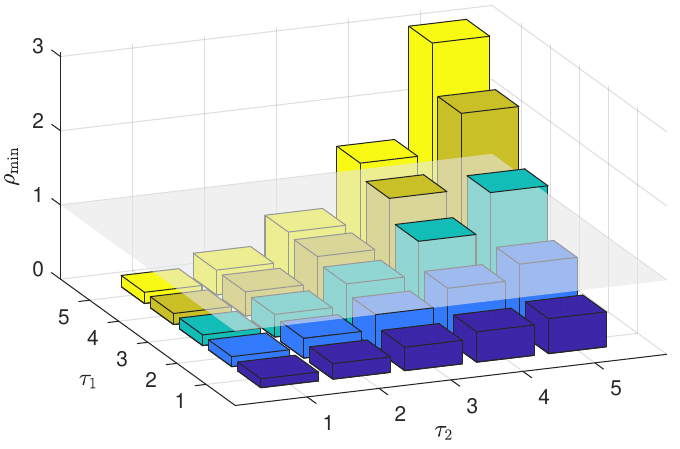}
  \caption{Stabilization radius $\rho_{\min}$ vs. input delays $\tau_1$ and $\tau_2$}\label{Fig:tau_vs_rho}
\end{figure}

Suppose that the same individual communication channels are adopted with $p_0 = 0.4,\,\alpha=\frac{2}{3}$, and $p_1 = 0.3$.
Also suppose that $\lambda = 1.5$, and $|s_1|<1$ and $|s_2|<1$, i.e., the minimum phase case.
Then the stabilization radius $\rho_{\min}$ with respect to the input delays $\tau_1$ and $\tau_2$ of the plant, is illustrated in Fig. \ref{Fig:tau_vs_rho}. 

\vspace{0.2cm}

It follows from Theorem \ref{Thm:Stabilizability_condition_delay} that the NCS is unstabilizable if $\tau_1>3$ and $\tau_2>4$.
It is interesting that, in contrast to SISO plant wherein the stabilization radius increases exponentially with the input delay of the plant (see \cite[Fig. 5]{Su2021meansquare}), by virtue of two inputs, the stabilization radius of the MIMO plant may increase slowly with one input delay when another input delay is small.
Moreover, it is intuitive that, when one of the input delays is large (i.e., the relevant input may be considered disconnected), the stabilization radius may nearly reduce to the SISO case that it increases exponentially with the other input delay.

\subsection{Example 3}

\begin{figure}[!htb]
  \centering
  \includegraphics[width=7.5cm]{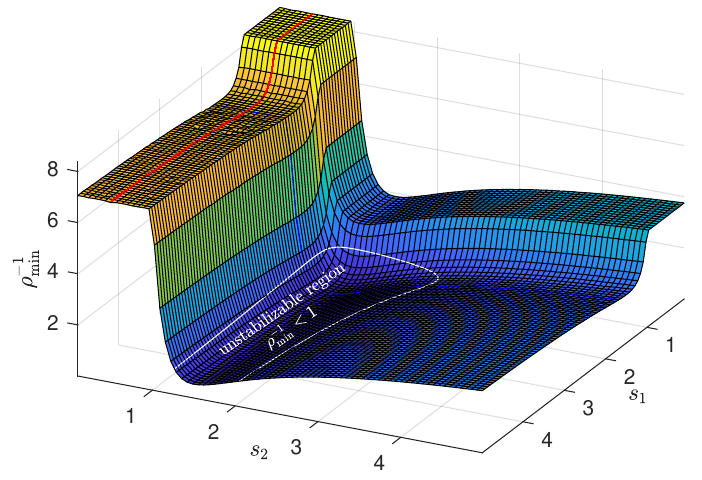}
  \caption{Inverse stabilization radius $\rho_{\min}^{-1}$ vs. zeros $s_1$ and $s_2$}\label{Fig:NMP_vs_rho}
\end{figure}

Suppose that $\tau_1=1$ and $\tau_2=1$, $\lambda=1.5$, and the channel is the same as in Example 2. 
Then the inverse stabilization radius $\rho_{\min}^{-1}$ with respect to the zeros of the plant, namely $s_1$ and $s_2$, is illustrated in Fig. \ref{Fig:NMP_vs_rho}.
One may realize that this is a generalization of \cite[Fig. 5]{Qi2017Control}. 

\section{Conclusion}
\label{Sec:Conclusion}

This work studied the mean-square stabilization for LTI MIMO systems cascaded by a linear stochastic system.
The linear stochastic system models a class of correlated stochastic uncertainties, such as channel uncertainties induced by packet loss and random transmission delays in networked systems.
The proposed mean-square stabilizability condition, both necessary and sufficient, for the closed-loop system revealed a fundamental limit imposed by the properties of the system (namely, the unstable poles, NMP zeros, and input delays) and the coefficient of frequency variation of the connected linear stochastic system.
Since the stabilizability was studied in the mean-square sense, the results only require the mean (nominal part) and spectral properties of the uncertainty, which implies that the results as well as analysis method in this work can be directly applied to uncertainties with similar simple frequency-domain properties.

\appendices

%

\section{Proof of \Cref{Thm:MS_stable_condition}}
\label{Appendix:MS_stability}

\hlt{Denote $G_{i_1i_2} $ the element located in the $i_1$th row and the $i_2$th column of $G$.}
To see the boundedness of the variances of $e(k)$ and $u(k)$, it suffices to show $\pow{e_i}$ and $\pow{u_i}$ are finite.
\hlt{Since the input $\{v(k)\}$ is a white noise sequence, Section 4 of \cite{Lu2022} shows that the signals in the closed loop are quasi-stationary. Moreover,} $S_u(e^{j\theta})\!=\!G(e^{j\theta})S_e(e^{j\theta})G^*(e^{j\theta})$.
Since $e = v-d$ and $v$ and $d$ are uncorrelated, $S_e(e^{j\theta})=S_v(e^{j\theta})+S_d(e^{j\theta})$.
Consequently, combining \eqref{Equ:S_d_MIMO} yields
\begin{align*}
S_u(e^{j\theta}) &= G(e^{j\theta})S_v(e^{j\theta})G^*(e^{j\theta}) \\
&\hspace{-1cm}+ G(e^{j\theta})S(e^{j\theta}) \diag\{ \|u_1\|_{\cal P}^2,\cdots,\|u_m\|_{\cal P}^2\} G^*(e^{j\theta}).
\end{align*}
By $\|u_i\|_{\cal P}^2 = \frac{1}{2\pi}\int_{-\pi}^{\pi}S_{u_i}(e^{j\theta})d\theta$, 
it is straightforward that
\begin{align*}
\begin{bmatrix}
\begin{smallmatrix}
\pow{u_1}^2\\
 \vdots \\
\pow{u_m}^2
\end{smallmatrix}
\end{bmatrix} &= \begin{bmatrix}\begin{smallmatrix}
{\left\| {{G_{11}}} \right\|_2^2}& \cdots &{\left\| {{G_{1m}}} \right\|_2^2}\\
 \vdots &{}& \vdots \\
{\left\| {{G_{m1}}} \right\|_2^2}& \cdots &{\left\| {{G_{mm}}} \right\|_2^2}
\end{smallmatrix}\end{bmatrix}\begin{bmatrix}\begin{smallmatrix}
               \sigma_{v_1}^2 \\
               \vdots \\
               \sigma_{v_m}^2
             \end{smallmatrix}\end{bmatrix} \\
&+\begin{bmatrix}\begin{smallmatrix}
{\left\| {{G_{11}}{\Phi _1}} \right\|_2^2}& \cdots &{\left\| {{G_{1m}}{\Phi _m}} \right\|_2^2}\\
 \vdots &{}& \vdots \\
{\left\| {{G_{m1}}{\Phi _1}} \right\|_2^2}& \cdots &{\left\| {{G_{mm}}{\Phi _m}} \right\|_2^2}
\end{smallmatrix}\end{bmatrix}  \begin{bmatrix}\begin{smallmatrix}
\pow{u_1}^2\\
 \vdots \\
\pow{u_m}^2
\end{smallmatrix}\end{bmatrix}.
\end{align*}
Then a necessary and sufficient condition for the existence of finite and unique $\pow{u_i}^2$ is that the condition \eqref{Equ:MS_stable_condition} holds, noting that $G_{i_1 i_2}\Phi_{i_2} = T_{i_1 i_2}W_{i_2}$.
Meanwhile, with the fact that $\pow{e_i}^2 = \sigma_{v_i}^2 + \pow{d_i}^2$, the proof is established.

\section{Proof of \Cref{Lma:M-f}}
\label{Appendix:M-f}

It follows from \eqref{Equ:Mgii_realization} that
$f_{\tau,i} \!=\! \hat{D}e_i \!+\! \sum_{j=1}^{\tau-1}\! \hat{C} \hat{A}^{j-1}\hat{B} e_i z^{-j}$.
Since $\mgi^{-1}\!=\! \hat{D} \!-\! \sum_{j=0}^{\infty} \!\hat{C} \hat{A}^{-j-1}\hat{B} z^j$, it holds $(f_{\tau,i}\!-\!\mgi^{-1}e_i)z^{\tau}\!= \!\sum_{j=1}^{\infty}\! \hat{C} {\hat{A}}^{\tau-1-j}\hat{B}e_i z^{j}$.
Let $\hat{G}(z) \!=\! \sum_{j=0}^{\infty} g_{j}z^{-j}$ be the Laurent series of $\hat{G}$.
Then
\begin{align*}
(f_{\tau,i}-\mgi^{-1}e_i)z^{\tau} \hat{G} 
&= \sum_{k=1}^{\infty} \sum_{j=0}^{\infty} \hat{C}\hat{A}^{\tau-1-j-k}\hat{B}e_i g_{j}z^k + \bar{Z}_{2i} \\
&=  \hat{C} (zI-\hat{A})^{-1} \hat{A}^{\tau-1} \hat{G}(\hat{A}) \hat{B}e_i z + \bar{Z}_{2i} 
\end{align*}
where $\bar{Z}_{2i} \!\in \!\mathcal{H}_{2}$.
Since $\sum_{k=1}^{\infty} \hat{A}^{\tau\!-\!1\!-\!k} z^k \!=\! (zI\!-\!\hat{A})^{-1} \hat{A}^{\tau-1}$, \eqref{Equ:Z1i_proj} holds and $\norm{Z_{1i}}{2}^2 \!=\! e_i^* \hat{B}^*\hat{G}^*(\hat{A})\hat{A}^{*^{\tau-1}} X \hat{A}^{\tau-1}\hat{G}(\hat{A})\hat{B}e_i$,  
where $X = \sum_{k=1}^{\infty} \big( \hat{A}^{-*k} \hat{C}^* \hat{C} \hat{A}^{-k} \big)$ is
exactly the solution to the Lyapunov equation $X \!=\! \hat{A}^{\!-*} X \hat{A}^{\!-1} \!+\! \hat{A}^{\!-*}\hat{C}^*\hat{C}\hat{A}^{\!-1}$, which is, by observing \eqref{Equ:Mgii_realization} and \eqref{Equ:M_Gamma_in}, equivalent to the DARE \eqref{Equ:X}.

\section{Proof of \Cref{Cor:Stabilizability_nonminimum_triangular}}
\label{Appendix:Stabilizability_triangular}

Let $PH = NM^{-1}$ be an upper triangular coprime factorization such that the diagonal elements $M_{ii},i=1,\cdots,m$ of $M$ are inner functions.
Let $\Ges = \diag\{1,\epsilon^{-1},\cdots,\epsilon^{-m+1}\}$ and $\mge = \Ges M \Gesi $.
Then it holds that $ \lim_{\epsilon \to 0}\mge = \diag\{M_{11},M_{22},\cdots,M_{mm}\}$,
which is inner.
It is clear that
\begin{align*}
\inf_{\Gamma}\inf_{Q} \rho(J(Q))
&\le \lim_{\epsilon \to 0} \inf_{Q} \| \Ges (U+MQ)\nt W \Gesi  e_i \|_2^2 \\
&= \| Z_{1}e_i\|_2^2 + \frac{|s_i|^2 -1}{|s_i|^2}\| Z_{1}(s_i)e_i\|_2^2
\end{align*}
where the last equality follows from \eqref{Equ:Ji_Q_NMP} with $Z_{1}$ being the $\rh{2}^\perp$ part of $(M_{in0}^{-1}-M_{in0}^{-1}(\infty))W z$.
Thus,
\begin{align*}
\rho_{\min}&\le \max_{i} \Big(\| Z_{1i}\|_2^2 + \frac{|s_i|^2 -1}{|s_i|^2}\| Z_{1i}(s_i)\|_2^2 \Big)
\end{align*}
where $Z_{1i}$ is the $\rh{2}^\perp$ part of $(M_{ii}^{-1}-M_{ii}^{-1}(\infty))W_i z$.
Note that from \cite{Zhou1995} there exists a balanced realization of $M_{ii}$ such that $\begin{bmatrix}\begin{smallmatrix}
\aii  &  \bii  \\
\cii  &  \dii
\end{smallmatrix}\end{bmatrix}
\begin{bmatrix}\begin{smallmatrix}
\aii^*  &  \cii^*  \\
\bii^*  &  \dii^*
\end{smallmatrix}\end{bmatrix}
=
\begin{bmatrix}\begin{smallmatrix}
I  &  0  \\
0  &  1
\end{smallmatrix}\end{bmatrix}$.
If this is the case, $\hat{A} = \aii^{-*}$ and the solution to the DARE \eqref{Equ:X} corresponding to $M_{ii}$ is $X=I$ provided that $\Gamma = 1$.
Then, together with the proof of \Cref{Thm:Rho_min_nonminimum}, the proof is established.

\bibliographystyle{IEEEtran}
\bibliography{mimo}

\begin{thebibliography}{10}
\providecommand{\url}[1]{#1}
\csname url@rmstyle\endcsname
\providecommand{\newblock}{\relax}
\providecommand{\bibinfo}[2]{#2}
\providecommand\BIBentrySTDinterwordspacing{\spaceskip=0pt\relax}
\providecommand\BIBentryALTinterwordstretchfactor{4}
\providecommand\BIBentryALTinterwordspacing{\spaceskip=\fontdimen2\font plus
\BIBentryALTinterwordstretchfactor\fontdimen3\font minus
  \fontdimen4\font\relax}
\providecommand\BIBforeignlanguage[2]{{%
\expandafter\ifx\csname l@#1\endcsname\relax
\typeout{** WARNING: IEEEtran.bst: No hyphenation pattern has been}%
\typeout{** loaded for the language `#1'. Using the pattern for}%
\typeout{** the default language instead.}%
\else
\language=\csname l@#1\endcsname
\fi
#2}}

\bibitem{Hespanha2007}
J.~P. Hespanha, P.~Naghshtabrizi, and Y.~Xu, ``A survey of recent results in
  networked control systems,'' \emph{Proceedings of the IEEE}, vol.~95, no.~1,
  pp. 138--162, Jan 2007.

\bibitem{ZHANG2017Survey}
D.~Zhang, P.~Shi, Q.-G. Wang, and L.~Yu, ``Analysis and synthesis of networked
  control systems: A survey of recent advances and challenges,'' \emph{ISA
  Transactions}, vol.~66, pp. 376 -- 392, 2017.

\bibitem{Park2018}
P.~Park, S.~C. Ergen, C.~Fischione, C.~Lu, and K.~H. Johansson, ``Wireless
  network design for control systems: A survey,'' \emph{IEEE Communications
  Surveys Tutorials}, vol.~20, no.~2, pp. 978--1013, 2018.

\bibitem{Willems1971Frequency}
J.~{Willems} and G.~{Blankenship}, ``Frequency domain stability criteria for
  stochastic systems,'' \emph{IEEE Transactions on Automatic Control}, vol.~16,
  no.~4, pp. 292--299, 1971.

\bibitem{Lu2002MeanSquare}
J.~Lu and R.~E. Skelton, ``Mean-square small gain theorem for stochastic
  control: discrete-time case,'' \emph{IEEE Transactions on Automatic Control},
  vol.~47, no.~3, pp. 490--494, Mar. 2002.

\bibitem{Qi2017Control}
T.~Qi, J.~Chen, W.~Su, and M.~Fu, ``Control under stochastic multiplicative
  uncertainties: Part {I}, fundamental conditions of stabilizability,''
  \emph{IEEE Transactions on Automatic Control}, vol.~62, no.~3, pp.
  1269--1284, Mar. 2017.

\bibitem{Elia2005Remote}
N.~Elia, ``Remote stabilization over fading channels,'' \emph{Systems \&
  Control Letters}, vol.~54, no.~3, pp. 237--249, 2005.

\bibitem{Xiao2009Mean}
N.~Xiao, L.~Xie, and L.~Qiu, ``Feedback stabilization of discrete-time
  networked systems over fading channels,'' \emph{IEEE Transactions on
  Automatic Control}, vol.~57, no.~9, pp. 2176--2189, 2012.

\bibitem{GonzalezChen2019}
R.~A. Gonz\'alez, F.~J. Vargas, and J.~Chen, ``Mean square stabilization over
  snr-constrained channels with colored and spatially correlated additive
  noises,'' \emph{IEEE Transactions on Automatic Control}, vol.~64, no.~11, pp.
  4825--4832, Nov 2019.

\bibitem{Zhang2020}
X.-M. Zhang, Q.-L. Han, X.~Ge, D.~Ding, L.~Ding, D.~Yue, and C.~Peng,
  ``Networked control systems: a survey of trends and techniques,''
  \emph{IEEE/CAA Journal of Automatica Sinica}, vol.~7, no.~1, pp. 1--17, Jan.
  2020.

\bibitem{Goldsmith2005Wireless}
A.~Goldsmith, \emph{Wireless Communications}.\hskip 1em plus 0.5em minus
  0.4em\relax Cambridge University Press, 2005.

\bibitem{QuevedoJ_TAC2014}
D.~Quevedo and J.~Jurado, ``Stability of sequence-based control with random
  delays and dropouts,'' \emph{IEEE Transactions on Automatic Control},
  vol.~59, no.~5, pp. 1296--1302, May 2014.

\bibitem{Su2017mean-square}
W.~Su, J.~Lu, and J.~Li, ``Mean-square stabilizability of a siso linear
  feedback system over a communication channel with random delay,'' in
  \emph{Proceedings of Chinese Automation Congress}, 2017, pp. 6965--6970.

\bibitem{Su2021meansquare}
\BIBentryALTinterwordspacing
W.~Su, J.~Li, and J.~Lu, ``Mean-square input-output stability and
  stabilizability of a networked control system with random channel induced
  delays,'' 2021, arXiv:2108.12795. [Online]. Available:
  \url{https://doi.org/10.48550/arXiv.2108.12795}
\BIBentrySTDinterwordspacing

\bibitem{Li2021Stabilization}
L.~Li, H.~Zhang, and Y.~Wang, ``Stabilization and optimal control of
  discrete-time systems with multiplicative noise and multiple input delays,''
  \emph{Systems \& Control Letters}, vol. 147, p. 104833, 2021.

\bibitem{Dipankar2022Optimal}
D.~Maity, M.~H. Mamduhi, S.~Hirche, and K.~H. Johansson, ``Optimal {LQG}
  control of networked systems under traffic-correlated delay and dropout,''
  \emph{IEEE Control Systems Letters}, vol.~6, pp. 1280--1285, 2022.

\bibitem{Xu2023}
J.~Xu, H.~Zhang, G.~Gu, and Y.~Tang, ``Stabilization of discrete-time systems
  with random delays and packet drops: a predictor-like controller,''
  \emph{SIAM Journal on Control and Optimization}, vol.~61, no.~3, pp.
  1737--1759, 2023.

\bibitem{Chen2024}
J.~Chen, T.~Qi, Y.~Ding, H.~Peng, J.~Chen, and S.~Hara, ``Mean-square stability
  radii for stochastic robustness analysis: A frequency-domain approach,''
  \emph{IEEE Transactions on Automatic Control}, pp. 1--16, Feb. 2024, {Early
  Access}.

\bibitem{Li2016Stabilization}
L.~Li and H.~Zhang, ``Stabilization of discrete-time systems with
  multiplicative noise and multiple delays in the control variable,''
  \emph{SIAM Journal on Control and Optimization}, vol.~54, no.~2, pp.
  894--917, 2016.

\bibitem{Zhou1995}
K.~Zhou, J.~C. Doyle, and K.~Glover, \emph{Robust and Optimal Control}.\hskip
  1em plus 0.5em minus 0.4em\relax Pearson, 1995.

\bibitem{Stewart2009Probability}
W.~J. Stewart, \emph{Probability, Markov Chains, Queues, and Simulation: The
  Mathematical Basis of Performance Modeling}.\hskip 1em plus 0.5em minus
  0.4em\relax USA: Princeton University Press, 2009.

\bibitem{Horn1986Matrix}
R.~A. Horn and C.~R. Johnson, \emph{Matrix Analysis}.\hskip 1em plus 0.5em
  minus 0.4em\relax New York, NY, USA: Cambridge University Press, 1986.

\bibitem{Lin1996Inner}
Z.~Lin, B.~M. Chen, A.~Saberi, and Y.~Shamash, ``Inner-outer factorization of
  discrete-time transfer function matrices,'' \emph{IEEE Transactions on
  Circuits and Systems I: Fundamental Theory and Applications}, vol.~43,
  no.~11, pp. 941--945, 1996.

\bibitem{Boyd1994Linear}
S.~Boyd, L.~{El~{G}haoui}, E.~Feron, and V.~Balakrishnan, \emph{Linear Matrix
  Inequalities in System and Control Theory}, ser. Studies in Applied
  Mathematics.\hskip 1em plus 0.5em minus 0.4em\relax Philadelphia, PA: {SIAM},
  June 1994, vol.~15.

\bibitem{Lu2019Mean-square}
J.~Lu, W.~Su, Y.~Wu, M.~Fu, and J.~Chen, ``Mean-square stabilizability via
  output feedback for a non-minimum phase networked feedback system,''
  \emph{Automatica}, vol. 105, pp. 142--148, 2019.

\bibitem{Lu2022}
\BIBentryALTinterwordspacing
J.~Lu, J.~Li, and W.~Su, ``Mean-square stability of linear systems over
  channels with random transmission delays,'' 2022. [Online]. Available:
  \url{https://doi.org/10.48550/arXiv.2204.13083}
\BIBentrySTDinterwordspacing

\end{thebibliography}

\end{document}